\documentclass [aps,amsmath,amssymb,superscriptaddress,preprint]{revtex4}
\usepackage{graphicx}
\usepackage{xcolor}

\def\gsim{\mathrel{\rlap{\lower4pt\hbox{\hskip1pt$\sim$}}
    \raise1pt\hbox{$>$}}}       

\begin{document}

\title{Fundamental Limit on Angular Measurements and Rotations from Quantum Mechanics and General Relativity}  

\author{Xavier Calmet} \email{x.calmet@sussex.ac.uk}
\affiliation{Department of Physics and Astronomy,\\
University of Sussex, Brighton, BN1 9QH, United Kingdom}

\author{Stephen~D.~H.~Hsu} \email{hsusteve@gmail.com}
\affiliation{Department of Physics and Astronomy\\ Michigan State University \\}

\begin{abstract}
\bigskip
We show that the precision of an angular measurement or rotation (e.g., on the orientation of a qubit or spin state) is limited by fundamental constraints arising from quantum mechanics and general relativity (gravitational collapse). The limiting precision is $r^{-1}$ in Planck units, where $r$ is the physical extent of the (possibly macroscopic) device used to manipulate the spin state. This fundamental limitation means that spin states $S_1$ and $S_2$ cannot be experimentally distinguished from each other if they differ by a sufficiently small rotation. Experiments cannot exclude the possibility that the space of quantum state vectors (i.e., Hilbert space) is fundamentally discrete, rather than continuous. We discuss the implications for finitism: does physics require infinity or a continuum?
\end{abstract}


\maketitle

\date{today}


\section{Introduction: Fundamental limits on measurement}

Gedanken experiments can reveal fundamental limitations on measurements or other experimental procedures that arise from the laws of physics, see e.g. \cite{Garay:1994en,Mead:1964zz,Padmanabhan:1987au,Baez:2002ra,Ng:1999hm,Hossenfelder:2012jw,Gibbons:2002iv}. The best known example of this is the Heisenberg uncertainty principle, which is motivated by elementary considerations of particle scattering. Heuristic arguments suggested that localization of a particle in position space would inevitable contribute to uncertainty in its momentum, and vice versa. It was only later that the formal theory of quantum mechanics incorporated this uncertainty in the form of operator commutation relations $[x, p] = i \hbar$. 

More recently, it has been shown that discreteness of space-time on length scales smaller than the Planck length cannot be detected due to limitations on measuring devices which arise from quantum mechanics and general relativity \cite{Calmet:2004mp,Calmet:2005mh,Calmet:2007vb}. This suggests, but does not prove, that models of quantum gravity that are consistent with what is currently known about low energy (long distance) physics will incorporate minimal length in some fundamental way. Examples are models which incorporate a minimal length via a generalized uncertainty principle \cite{Maggiore:1993kv,Kempf:1994su,Scardigli:1999jh,Scardigli:2003kr} or even string theory.

In this letter we deduce the existence of minimal rotations which are analogous to minimal length. Specifically, we deduce limits on the precision with which a rotation can be applied (e.g., to a microscopic state), or a device can measure spin orientation. These results suggest a fundamental discreteness in the structure of Hilbert space itself \cite{BHZ1,BHZ2}. 

The result described here was first obtained in \cite{BHZ1} using minimal length (the Planck length) as an input assumption. Consider the rotation of a macroscopic device of size $r$. If the angle of rotation is sufficiently small no part of the device is displaced by more than the minimal length, and the device is not distinguishable from its unrotated configuration. Thus rotations smaller than $r^{-1}$ in Planck units cannot be realized and measurements with better than this precision cannot be performed.

Below we give a more complete derivation of the result: we consider the angular displacement operator $\varphi (t) - \varphi (0)$ and examine limits on related experimental procedures. This is analogous to the approach used in \cite{Calmet:2004mp} to deduce minimal length.

\section{Angular measurements and rotations}

In this section, we show that quantum mechanics and general relativity considered simultaneously imply the existence of a bound on the precision of the measurement of an angle: i.e.,
no operational procedure exists which can measure an angle less than this fundamental angle. The key ingredients used to reach this conclusion are the uncertainty principle from quantum
mechanics, and gravitational collapse from general relativity in the form of the hoop conjecture.

A dynamical condition for gravitational collapse is given by the Hoop
Conjecture \cite{hoop}: if an amount of energy $E$ is confined at any instant to a ball of size $R$, where $R < G_N E$ ($G_N$ is Newton's constant),
then that region will eventually evolve into a black hole. Recent theoretical results \cite{Penrose74,D'Eath:1992hb,D'Eath:1992hd,D'Eath:1992qu,Eardley:2002re,Hsu:2002bd,Calmet:2014dea} support the Hoop Conjecture : even the scattering of extremely high energy particles cannot avoid black hole formation if their impact parameter is less than of order $R$ given above. Henceforth, we use natural units where $\hbar, c$ and Newton's constant (or $l_P$) are unity. We also sometimes neglect numerical factors of order one.


Let us consider the Hamiltonian for a free particle on a circle of radius $r$. The Hamiltonian is given by 
\begin{eqnarray}
H=\frac{p_\varphi^2}{2 m r^2},
\end{eqnarray}
where $p_\varphi$ is the angular momentum. The angular position of the particle is given by the classical equation of motion
\begin{eqnarray}
\dot \varphi=\frac{\partial H}{\partial p_\varphi}=\omega,
\end{eqnarray}
which admits the solution
\begin{eqnarray}
\varphi(t)= \omega t + \varphi(0).
\end{eqnarray}

Passing to quantum mechanics, we can calculate the commutator of $\varphi(t)$ and $\varphi(0)$ and obtain
\begin{eqnarray}
[\varphi(0),\varphi(t)]=i \frac{t}{m r^2},
\end{eqnarray}
where $t$ is the time between the two measurements.
From this we obtain
\begin{eqnarray}
|\Delta \varphi(0)| |\Delta \varphi(t)|\ge \frac{ t}{2 m r^2}.
\end{eqnarray}
As measuring an angle requires two measurements, it is limited by the greater of  
$\Delta \varphi(0)$ and $\Delta \varphi(t)$, which is at least $( t / 2 m r^2 )^{1/2}$. At first sight, it appears that one could make this difference arbitrarily small by making $m$ very large. However, in order to avoid gravitational collapse, the size $r$ of the object must scale proportionally to $m$ so that $r > m$ (the inequality holds up to factors of order one). By causality $r$ cannot exceed $t$. We thus find
\begin{eqnarray}
|\Delta \varphi(t)|\ge \frac{l_P}{\sqrt{2} r},
\end{eqnarray}
where, leaving natural units momentarily, $l_P$ is the Planck length given by $\sqrt{\hbar G_N/c^3}\sim 1.6 \times 10^{-35}$ m. The uncertainty in the measurement of $\varphi$ can be reduced by making $r$ large, but it cannot approach zero without taking the rotational inertia of the device to infinity. Of course, more practical limitations resulting from material properties (e.g., causal bounds on rigidity) will intervene before this limit can be taken.

We neglected interactions between the experimental apparatus and the region of the universe outside the causal region of the measurement. The possibility of interactions, and a more complicated Hamiltonian, were considered in \cite{Calmet:2004mp} and shown not to alter the conclusions.


Our result can be interpreted as a proof of the existence of a minimal angle in full analogy to the minimal length related to the Planck mass. This minimal angle could be called the Planck angle. Basically, it implies that no operational procedure can exclude the quantization of space-time for distances or angles less than the Planck length or the Planck angle. We emphasize that our result does not rely on a discretization of space-time; the angular evolution of the object on the circle is given by standard quantum mechanics on a continuous space-time.   However, given the existence of a minimal length, one could speculate that space-time has a short distance (high momentum) cutoff, and thus physics in a bounded region of space-time is described by a finite dimensional Hilbert space. We have shown that no experiment could rule this out. 

We note that canonical commutation relations such as $[ x, p ] = i\hbar$ cannot be realized in a finite dimensional Hilbert space. This is easy to verify by taking the trace of both sides of the equation: the left hand side is traceless but the right hand side has trace proportional to the dimensionality \cite{finite}. However, this is primarily a technical issue because finite dimensional quantum systems continue to obey the usual uncertainty relations. For example, a wavepacket state realized in a discrete and finite setting (e.g., a space-time lattice with finite volume) cannot reduce its position uncertainty $\Delta x$ arbitrarily without increasing the corresponding momentum uncertainty $\Delta p$, and vice versa. It is sometimes argued (erroneously) that quantum mechanics must have an infinite dimensional Hilbert space because of the position-momentum commutation relation. However in fact what is really known about quantum physics from direct observation is not the commutation relation itself but the uncertainty relation that it encodes. The uncertainty relation persists in a finite dimensional version of quantum mechanics (e.g., the Schr\"odinger evolution of states on a discrete space-time lattice).

\section{Superpositions and Spin States}

The results of the previous section limit the precision with which we can measure or manipulate the state of a single qubit -- the orientation of a spin:
\begin{equation}
\vert \psi \rangle = \cos \theta \, \vert + \rangle ~+~ e^{i \phi} \sin \theta \, \vert - \rangle~~~.
\label{qubit}
\end{equation}
Limits on the precision of $( \theta, \phi )$ are actually limits on the precision of Hilbert space (or state vector) coefficients. These limits, together with the Planck length short distance cutoff, imply that no experiment can exclude the possibility that Hilbert space is discrete and finite dimensional, i.e.,  
\begin{equation}
\label{psisum}
\vert \psi \rangle = \sum c_n \, \vert n \rangle ~~~
\end{equation}
where  1. the values of the coefficients $c_n$ are only defined to some finite accuracy -- they are {\it not} continuous complex parameters, and 2. the sum itself is finite.

Physicists who simulate quantum phenomena on classical computers are already familiar with properties 1 and 2. What we describe above as fundamental consequences for quantum mechanics arising from gravity are approximations made out of necessity in everyday computation. 
Any quantum phenomenon that can, in principle, be reproduced to satisfactory approximation in computer simulation is entirely consistent with properties 1 and 2. 
\bigskip

We can also pursue a quantum information approach to these questions.
Consider an experiment which takes place in a space-time region of extent $r$. Given the short distance cutoff at the Planck length, $l_p$, the number of degrees of freedom relevant to the experiment is itself finite. For a given $r$, the number of distinct configurations of the experimental apparatus (i.e., the number of distinct quantum operators represented by the possible measurements) is bounded above. Thus the number of distinct spin orientations (qubit states which are eigenstates of the measurement operator) that can be resolved is also bounded above. Physics can therefore be described by a discretized Hilbert space in which the angles $( \theta, \phi )$ are discrete and take on only a finite (but presumably very large) number of values. 

Holography (another aspect of quantum gravity) provides a stronger bound on the scale of discreteness: the total entropy of the measurement apparatus is bounded above (i.e., limiting its configurations and accuracy of read out) by the boundary area rather than the volume of the region in Planck units.

We can illustrate the connection to holography \cite{BHZ1} using a composite state built from many qubits:
\begin{equation}
\Psi = \psi_1 \otimes \psi_2 \otimes \psi_3 \otimes \cdots \otimes \psi_n~.
\end{equation}    
Assume a fundamental uncertainty $\epsilon$ in the state of each of the qubits, so that $\psi$ and $\psi'$ are indistinguishable when $\vert \psi - \psi' \vert < \epsilon$. Now consider the composite state
\begin{equation}
\Psi' = \psi'_1 \otimes \psi'_2 \otimes \psi'_3 \otimes \cdots \otimes \psi'_n~~~.
\end{equation}  
If uncertainties $\epsilon$ for each qubit are independent, the resulting uncertainty in the composite state $\Psi$ is
\begin{equation}
\vert \Psi - \Psi' \vert^2 \sim n \epsilon^2 
\end{equation}
and requiring that this be less than unity implies the holographic bound
\begin{equation}
n ~<~ \epsilon^{-2} ~\sim~ r^2 
\end{equation}
where $r$ is the size of the system. In other words, the requirement of a small aggregate uncertainty in the composite state $\Psi$ due to the individual qubit uncertainties $\epsilon$ is equivalent to the holographic upper bound on the entropy or number of degrees of freedom $n$ in a finite region of space of size $r$.

To summarize, the observation that only a finite amount of quantum information can be encoded in a finite region provides an upper limit on the precision of a measurement conducted in that region. However, we can go further: the measured state of a qubit can only be determined to some limited precision, and this is consistent with a finite (rather than infinite, as is usually assumed) set of possible orientations $( \theta, \phi )$ {\it for each qubit in the universe}.

\section{Finitism: Does Physics require a Continuum?}

Our intuitions about the existence and nature of a continuum arise from perceptions of space and time \cite{Weyl}. But the existence of a fundamental Planck length suggests that space-time may not be a continuum. In that case, our intuitions originate from something (an idealization) that is not actually realized in Nature.

Quantum mechanics is formulated using continuous structures such as Hilbert space and a smoothly varying wavefunction, incorporating complex numbers of arbitrary precision. However beautiful these structures may be, it is possible that they are idealizations that do not exist in the physical world. The introduction of gravity limits the precision necessary to formulate a model of fundamental quantum physics. Indeed, any potential structure smaller than the Planck length or the minimal angle considered here cannot be observed by any device subject to quantum mechanics, general relativity, and causality. Our results suggest that quantum mechanics combined with gravity does not require a continuum, nor any concept of infinity.

It may come as a surprise to physicists that infinity and the continuum are even today the subject of debate in mathematics and the philosophy of mathematics. Some mathematicians, called {\it finitists}, accept only finite mathematical objects and procedures \cite{Ye}. The fact that physics does not require infinity or a continuum is an important empirical input to the debate over finitism. For example, a finitist might assert (contra the Platonist perspective adopted by many mathematicians) that human brains built from finite arrangements of atoms, and operating under natural laws (physics) that are finitistic, are unlikely to have trustworthy intuitions concerning abstract concepts such as the continuum. These facts about the brain and about physical laws stand in contrast to intuitive assumptions adopted by many mathematicians. For example, Weyl (Das Kontinuum \cite{Weyl,Feferman}) argues that our intuitions concerning the continuum originate in the mind's perception of the continuity of space-time.

There was a concerted effort beginning in the 20th century to place infinity and the continuum on a rigorous foundation using logic and set theory. As demonstrated by G\"odel, Hilbert's program of axiomatization using finitary methods (originally motivated, in part, by the continuum in analysis) could not succeed. Opinions are divided on modern approaches which are non-finitary. For example, the standard axioms of Zermelo-Fraenkel (ZFC) set theory applied to infinite sets lead to many counterintuitive results such as the Banach-Tarski Paradox: given any two solid objects, the cut pieces of either one can be reassembled into the other \cite{BT}. When examined closely all of the axioms of ZFC (e.g., Axiom of Choice) are intuitively obvious if applied to finite sets, with the exception of the Axiom of Infinity, which admits infinite sets. (Infinite sets are inexhaustible, so application of the Axiom of Choice leads to pathological results.) The Continuum Hypothesis, which proposes that there is no cardinality strictly between that of the integers and reals, has been shown to be independent (neither provable nor disprovable) in ZFC \cite{cohen}. Finitists assert that this illustrates how little control rigorous mathematics has on even the most fundamental properties of the continuum.

\bigskip

David Deutsch \cite{Deutsch}:
\smallskip

\begin{quote}
{\it The reason why we find it possible to construct, say, electronic calculators, and indeed why we can perform mental arithmetic, cannot be found in mathematics or logic. The reason is that the laws of physics ``happen to'' permit the existence of physical models for the operations of arithmetic such as addition, subtraction and multiplication.}
\end{quote}

\smallskip

This suggests the primacy of physical reality over mathematics, whereas usually the opposite assumption is made. From this perspective, the parts of mathematics which are simply models or abstractions of ``real'' physical things are most likely to be free of contradiction or misleading intuition. Aspects of mathematics which have no physical analog (e.g., infinite sets, the continuum) are prone to problems in formalization or mechanization. Physics -- i.e., models which can be compared to experimental observation, actual ``effective procedures'' -- does not ever require infinity, although it may be of some conceptual convenience. Hence it seems possible, and the finitists believe, that the Axiom of Infinity and its equivalents do not provide a sound foundation for mathematics.

\bigskip

{\it Acknowledgments:}
The work of X.C. is supported in part  by the Science and Technology Facilities Council (grants numbers ST/T00102X/1, ST/T006048/1 and ST/S002227/1). 

\end{document}